\documentclass[10pt,a4paper]{article}
\usepackage[ruled]{algorithm2e}
\usepackage{rotating}
\usepackage{amsmath,amssymb,amsbsy}
\usepackage{graphicx}
\usepackage{subfigure}
\usepackage{xspace}
\usepackage{wrapfig}

\SetAlFnt{\small}
\SetAlCapFnt{\small}
\SetAlCapNameFnt{\small}
\SetAlCapHSkip{0pt}
\IncMargin{-\parindent}

\title{Application of the Computer Capacity to the Analysis of Processors Evolution}
\author{BORIS RYABKO${}^1$ and 
ANTON RAKITSKIY${}^2$
}
\begin{document}
\maketitle

\begin{abstract}
 The notion of computer capacity was proposed in 2012, and this quantity has been estimated for computers of different kinds. 
 In this paper we  show that, when designing new processors, the manufacturers change the parameters that affect the computer capacity.
This allows us to predict the values of parameters of future processors. As the main example 
we use Intel processors,  due to 
the accessibility of detailed description of all their technical characteristics.
\end{abstract}

%


\footnotetext[1] {Institute of Computational Technologies of the Siberian Branch of the RAS, Novosibirsk State University}
\footnotetext[2] {Siberian State University of Telecommunication and Information Sciences, Institute of Computational 
Technologies of the Siberian Branch of the RAS}


%

\noindent
This work was supported by Russian Foundation for Basic Research (grant 15-07-01851) and by Grant of Leading Scientific Schools (SS-7214.2016.9)


\newpage
\setcounter{page}{1}
\section{Introduction}
A  theoretical approach to estimating computer capacity was suggested in 2012 \cite{perf_eval}. This approach uses only the description of the 
computer architecture investigated. The latter include the set of instructions, features of their execution, sizes of all the memory types, etc. 
Thus, no experiments with a working model of the computer are necessary. This method of estimating the computer capacity was used  to evaluate  
the performance  of a large number of different Intel and AMD processors \cite{jcsc}. In \cite{jcsc} we show that this characteristic is consistent 
with the values obtained empirically  using benchmarks \cite{weicker1991detailed,overall,passmark} (some of the results are presented in the appendix).
This method was also used to estimate the performance of some supercomputers from TOP500 list \cite{red16}.

In this paper we apply the method of estimating the computer capacity to study the evolution of  processors. 
We  assume that 
processor manufacturers are interested in changing those characteristics of processors that  give the largest
improvement in performance. Therefore, they need to be able to estimate the influence of changing some characteristic of a processor on its performance. 
At present, manufacturers use  benchmarks to evaluate the performance, but this method is too complex and requires a working model of the  computer investigated.
Besides, benchmarks are inappropriate to use for the estimation of characteristics influence due to the necessity of building a working model of processor 
for each change.

Here we consider the evolution of Intel processors for the last 15 years. 
Each processor can be represented as a set of parameters 
and changing some of these parameters has a significant effect on its performance. Our investigation shows that in new processors the manufacturers usually 
increase the characteristics which affect the computer capacity the most. This, in turn, allows us to predict the direction of changes in the evolution of 
computers. 

First, we note that there are clear tendencies that can be traced in the development of processors for the last 15 years.
The first obvious tendency is the increase of the clock 
rate which causes the decrease of the task execution time. The second one is the widespread introduction of parallelism, in particular: by increasing the
number of processors in computer; by increasing the number of computing cores in processor; by introducing threads and pipelines, etc. It is clear that the effect of these parameters is huge, but it is also obvious. 

The emphasis in our work is on the quantitative estimation of the impact of the parameters whose role in the performance is not so obvious. 
These parameters include: sizes and access times to different kinds of memory (including registers, cache-memory etc.); the number of different instructions 
of certain type (instruction types are characterised by the number and kind of the operands).

For example, if we examine the Intel processor codenamed Wolfdale, we see that the increase of some parameters significantly affects performance, whereas
changing the rest of the parameters has almost no effect. This, if we increase the size of level-1 cache memory in Wolfdale the value of the computer
capacity almost does not change, but if the number of internal vector registers is increased, the growth of the computer capacity becomes perceptible. 
It turns out that in the processors of the succeeding microarchitectures (Sandy Bridge, Ivy Bridge etc.) these exact parameters were increased.  
This allows us to claim that using the suggested method may be useful for predicting which parameters will be changed in succeeding the future models.

\section{Computer Capacity}\label{sec1}

All the theory behind the computer capacity and its estimation was described in details in the previous work \cite{perf_eval}, so here we only present
the main definitions and a brief summary of the theory required in order to understand the results.
 
Let us consider a computer with the set of instructions $I$ and memory $M$. An instruction $x \in I$ is formed as the combination of its name and the
values of its operands (two instructions with the same names and different values of operands are both included in $I$). For example, 
instructions \emph{mov eax, ebx} and \emph{mov eax, ecx} are different and included in $I$ independently. A computer task $X$ is a finite sequence of 
instructions $X=x_1 x_2 \dots, x_i \in I$. It is important to note that if there is a loop in a task which is repeated $m$ times, the body of this loop is
included $m$ times in $X$. We denote $\tau(x)$ the execution time of instruction $x$. So the execution time of a computer task $X$ is given 
by $\tau(X) = \sum_{i=1}^{n}{\tau(x_i)}, \, X=x_1 x_2 \dots x_n$. Let consider the number of all possible computer tasks which execution times equal
to $T$ as $N(T)=|\{X:\tau(X)=T\}|$. 

Let, for example,  there be a processor which has exactly $N(1)$ different tasks with execution times equal to 1 hour. In this case we can say 
that it can execute $N(1)^2$ different tasks in 2 hours because if instruction sequences $X_1$ and $X_2$ are 1-hour tasks, 
the combined sequence $X_1 X_2$ is the 2-hour one
(we did not take into account the 2-hour tasks with instruction starts at the end of the first hour and finishes and the beginning of the second,
because the share of such sequences is negligible).
In this way, the considered processor has $\approx N(1)^k$ tasks with execution times $k$ hours. So we can see that the number of possible tasks grows 
exponentially as a function of time ($N(T)\approx2^{CT}$). Therefore, $C=\frac{\log{(N(T))}}{T}$ (or rather the limit of this value) is the adequate measure of the computer capacity. This limit is defined as follows:
\begin{equation}
\label{eq1}
C(I)=\lim_{T \to \infty}{\frac{\log{N(T)}}{T}}.
\end{equation}

The main question here is how to estimate the value of $C(I)$ from (\ref{eq1}). A direct calculation of the limit is impossible, but it is possible to calculate  $C(I)$ with a method from   combinatorial analysis. To do this,  consider the set of instructions $I$ as an alphabet and assume 
that all words (sequences of instructions) over this alphabet are permitted (can be executed). This assumption allows us to estimate an upper-bound of 
the computer capacity, because for any processor the set of its permissible tasks is the subset of all possible tasks. Here, all execution times are 
integers (this statement is valid for  most of the processors if the time unite is equal to the clock rate). 
A way to estimate  the capacity was suggested by C. Shannon \cite{shannon}, 
who showed that the considered upper-bound of the capacity $C(I)$ is equal to the logarithm of the largest
real solution $Y_0$ of the following characteristic equation:
\begin{equation}
\label{eqCC}
Y^{-\tau (x_1)}+Y^{-\tau (x_2)}+\dots +Y^{-\tau (x_n)}=1.
\end{equation}
In \cite{perf_eval} it was also shown that the computer capacity of a multi-core processing unit can be defined as the sum of capacities of the cores.

\section{The computer capacity of Intel processors}
The present work is based on the analysis of Intel processors produced over the last 15 years, because the information about these processors architectures
(with the full description of instruction set) is public and easily accessible. In our previous works we have shown that the computer capacity 
correlates well with the values of benchmarks \cite{phansalkar2005measuring, passmark} and can be used as a measure of computer performance \cite{jcsc}.

We have identified a list of processors to analyze: Pentium M (Dothan processor), Intel Core (Wolfdale), Ivy Bridge, Haswell and Skylake. In this paper we  analyze these processors in details, evaluate their computer capacity values and perform some investigations about the effect of parameters 
on performance. 
Pentium M and Wolfdale architecture differs strongly from Ivy Bridge, Haswell and Skylake, so the comparison of this architectures would be quite 
interesting. Skylake is considered here because it is the latest Intel microarchitecture with a  detailed description  published (at the time of the preparation of this paper). Here, we present the details of the calculation of the computer capacity for Intel Pentium M, Core, Ivy Bridge, Haswell 
and Skylake microarchitectures. The structures of the described processors pipelines are similar to each other, and the features of building 
the equation (\ref{eqCC}) for that structure are presented in \cite{red12_1,jcsc}. 

In table \ref{tabSum} we present the summary of technical characteristics and the values of the computer capacity for all the  processors chosen.
The following notation is used: L1,L2,L3 is the size of level-1, level-2 and level-3 caches, M is the size of RAM, $L1_{t},L2_{t},L3_{t},M_{t}$ is the memory 
latencies for level-1, -2, -3 and RAM respectively, c.c. stands for clock cycles, $R_i$ is the number of integer registers, $R_v$ is the number of vector 
(floating point) registers. The detailed description and the lists of instructions of the described microarchitectures are presented in \cite{agner1}. 
The characteristic equations of all the processors described and all the tools for evaluation of the computer capacity can be found in \cite{link}. 

\begin{table}

\centering

\caption{The summary of characteristics for selected processors}\label{tabSum}

\begin{tabular}{|l|c|c|c|c|c|}

\hline

Processor &Pentium M&Intel&Ivy&Haswell&Skylake\\

&&Core&Bridge&&\\

\hline

L1, KB& 32 & 64 & 64 & 64 & 64\\

\hline

L2, KB& 2048 & 6144 & 256 & 256 & 256\\

\hline

L3, MB& - & - & 8 & 8 & 8 \\

\hline

M, GB& 1 & 16 & 16 & 16 & 16\\

\hline

$L1_{time}$, c.c.& 3 & 3 & 4 & 4 & 4\\

\hline

$L2_{time}$, c.c.& 10 & 15 & 12 & 12 & 12\\

\hline

$L3_{time}$, c.c.& - & - & 30 & 36 & 42\\

\hline

$M_{time}$, c.c.& 70 & 24 & 30 & 36 & 42\\

\hline

$R_i$& 8 & 16 & 160 & 168 & 180\\

\hline

$R_v$& 8 & 16 & 144 & 168 & 168\\

\hline

Computer& 51.217 & 70.898 & 108.587 & 115.86 & 116.208\\

capacity,&&&&&\\

bits/c.c.&&&&&\\

\hline

\end{tabular}

\end{table}

\section{Analysis of Intel processors evolution}
\label{sec:2}

To analyze the evolution of Intel processors we divide them into 5 groups: (Pentium M, Intel Core), (Intel Core, Ivy Bridge), (Ivy Bridge, Haswell),
(Haswell, Skylake) and (Skylake and the prediction for its successor). The investigated characteristics are the following:  
\begin{itemize}
\item Physical characteristics of processor: the size of all memory types; the access time of all memory types; the number of different registers. 
Here, the memory types are the 1-level cache, the 2-level cache, the 3-level cache (if processor contains this level) and the main memory (RAM).
Registers are also of two types: integer and vector. 
\item The instruction set is another characteristic that affects the performance. It can be observed that the instruction set changes from one 
processor to another. Here, we isolate the fastest instructions from instruction set of each examined processor, and calculate their number. 
The fastest instruction is the one whose execution time equals 1 and does not have memory cells in the list of operands (access to the memory cell greatly 
increases the execution time of an instruction). We group all these instructions by the number of operands and present the results in table \ref{tabops} 
(the names of columns signify the number of operands in the investigated instructions: the number of instructions with single operand (second column),
with two operands and with three operands). In most cases, these operands are registers of different types. 
\end{itemize}

\begin{table}[!ht]
\centering
\caption{Number of different instructions}\label{tabops}
\begin{tabular}{|l|l|l|l|}
\hline number of operands
&1&2&3\\
\hline
Pentium M (Dothan)&53&91&-\\
\hline
Intel Core (Wolfdale)&67&328&-\\
\hline
Ivy Bridge&21&99&10\\
\hline
Haswell&31&113&44\\
\hline
Skylake&35&134&48\\
\hline
\end{tabular}
\end{table}

We examined the influence of changing the value of a single characteristic (from the physical group) and identified those of them which have a nonzero influence on the value of the computer capacity. We also examined all the possible pairs of different characteristics, but only one pair ($R_i$,$R_v$) was included because all other pairs either have no effect on the value of the computer capacity or their effect equals the effect of a single characteristic. The values were calculated at $\times 0.5, \times 2, \times 5, \times 10, \times 20$ of the original value. 

Next we try to add instructions in the instruction set and to show the influence of this addition on the value of the computer capacity. The number of instructions in the modified instruction set is obtained as the original value $\times 1.1, \times 1.25, \times 1.5, \times 2$. The last step is the combination of the two previous steps. Here, we want to show the influence of increasing the number of registers and adding instructions of a new type simultaneously. It is shown above that in the Ivy Bridge processors a new type of instructions appeared: instructions with three register operands, and at the same time the number of registers is increased almost 10 times in relation to Intel Core (Wolfdale) processors. So, in this part we add the instructions of a new type: for Pentium M and Intel Core we add instructions with 3 operands, for the remaining processors we add instructions with four operands. In the experiments we add 8, 16, 32 and 64 instructions. Increasing the number of registers is performed in the following way: $\times2, \times 5, \times 10$ of the original value. 
Following is the list of added instructions types in the second and third steps: 
\begin{enumerate}
\item cmd r 1 / cmd x 1 - the instruction with the name "cmd", with a single integer register operand (r) or vector register (x) and with the execution time equals to 1;
\item cmd r,r 1 / cmd x,x 1- the instruction with two integer or two vector register operands;
\item cmd r,r,r 1 / cmd x,x,x 1 - the instruction with three integer or three vector register operands, codenamed  cmd1 and cmd2 respectively in the last step.
\item cmd r,r,r,r 1 / cmd x,x,x,x 1 - the instruction with four integer or four vector register operands, codenamed  cmd3 and cmd4 respectively in the last step.
\end{enumerate}
All the results in the tables are presented in percent relative to the original value of the computer capacity. The characteristics with the same results are merged into a single row.

\subsection{Pentium M and Intel Core}
The results of the analysis of Pentium M are presented in Tables \ref{tabpm1},\ref{tabpm2} and \ref{tabpm3}, one for each step of investigation.  The first row is filled with 100 and it means that the characteristics from this row have no influence on the computer capacity. So the role of the size and the access time of cache-memory and RAM are insignificant for the computer capacity. In the following subsections we exclude from the tables those characteristics that do not affect the value of computer capacity. We can also observe that the increase in the number of instructions has no significant effect (for the instructions of existing types). However, there are some characteristics which change the value of the computer capacity by more than 1\%. Obviously, to increase the capacity, we need to increase the number of registers (integer or vector) and add some instructions of a new type. As we can see in Tables  \ref{tabSum} and \ref{tabops}, the number of registers was doubled in Intel Core, the size of memory of different kinds was also increased, but the access times are almost unchanged. Increase in the Intel Core computer capacity is also explained by the improvement of the throughput of its pipeline (it grew from 3 $\mu$ops per cycle to 4).

\begin{table}[!ht]
\centering
\caption{Pentium M step 1}
\label{tabpm1}
\begin{tabular}{|l|l|l|l|l|l|}
\hline
&0.5&2&5&10&20\\
\hline
$L1,L2,M,$&&&&&\\
$L1_{t},L2_{t},M_{t}$&100&100&100&100&100\\
\hline
$R_i$&99.92&100.24&101.58&104.95&111.91\\
\hline
$R_v$&99.78&100.78&104.75&111.87&121.88\\
\hline
$R_i$ \& $R_v$&99.72&101.03&105.97&113.95&124.39\\
\hline
\end{tabular}
\vskip1pt
\centering
\caption{Pentium M step 2}
\label{tabpm2}
\begin{tabular}{|l|l|l|l|l|}
\hline
&1.1&1.25&1.5&2\\
\hline
"cmd r 1",&&&&\\
"cmd x 1"&100.002&100.006&100.013&100.026\\ 
\hline
"cmd r,r 1",&&&&\\
"cmd x,x 1"&100.035&100.086&100.175&100.35\\
\hline
\end{tabular}
\vskip1pt
\caption{Pentium M step 3}
\label{tabpm3}
\begin{tabular}{|l|l|l|l|l|}
\hline
&8&16&32&64\\
\hline
$r \times 2$ cmd1, $x \times 2$ cmd2&102.644&103.999&106.191&109.372\\
\hline
$r \times 5$ cmd1, $x \times 5$ cmd2&114.767&118.985&123.885&129.219\\
\hline
$r \times 10$ cmd1, $x \times 10$ cmd2&130.027&135.229&140.739&146.417\\
\hline
\end{tabular}
\end{table}

\subsection{Intel Core and Ivy Bridge}

In tables \ref{tabcore1},\ref{tabcore2} and \ref{tabcore3} the results are close to those for the previous processor, except for the effect of adding the instructions with 2 operands. We noted that the 10-20 fold increase of the number of registers and adding the instructions with three operands gives the best effect on the value of computer capacity. Indeed, the manufacturer change exactly this characteristics in the succeeding processor. In Ivy bridge the number of integer and vector registers increased tenfold and 10 fast instructions with 3 register operands were added.

\begin{table}[!ht]
\centering
\caption{Intel Core step 1}
\label{tabcore1}
\begin{tabular}{|l|l|l|l|l|l|}
\hline
&0.5&2&5&10&20\\
\hline
$R_i$&99.73&100.79&104.36&110.67&119.84\\
\hline
$R_v$&97.48&105.81&118.08&128.87&140.01\\
\hline
$R_i$ \& $R_v$&97.18&106.3&119&129.91&141.09\\
\hline
\end{tabular}
\vskip1pt

\caption{Intel Core step 2}
\label{tabcore2}
\begin{tabular}{|l|l|l|l|l|}
\hline
&1.1&1.25&1.5&2\\
\hline
"cmd r 1",&&&&\\
"cmd x 1"&100.004&100.01&100.02&100.04\\ 
\hline
"cmd r,r 1",&&&&\\
"cmd x,x 1"&100.302&100.753&101.442&102.667\\
\hline
\end{tabular}
\vskip1pt
\caption{Intel Core step 3}
\label{tabcore3}

\begin{tabular}{|l|l|l|l|l|l|}
\hline
&8&16&32&64\\
\hline
$r \times 2$ cmd1, $x \times 2$ cmd2&109.908&112.401&115.849&120.124\\
\hline
$r \times 5$ cmd1, $x \times 5$ cmd2&127.474&131.534&136.253&141.391\\
\hline
$r \times 10$ cmd1, $x \times 10$ cmd2&142.739&147.493&152.651&158.04\\
\hline
\end{tabular}
\end{table}

\subsection{Ivy Bridge and Haswell}
The results for Ivy Bridge analysis are presented in tables \ref{tabivy1}, \ref{tabivy2} and \ref{tabivy3}. Here, we observe tendencies similar to those in the previous processors. It is interesting to observe that starting from Ivy Bridge the characteristics related to cache-memory and RAM are almost unchanged. We can also notice that starting with Ivy Bridge the effect of the number of commands with one register becomes insignificant and we exclude them from the tables of the following subsections. 
\begin{table}[!ht]
\centering
\caption{Ivy Bridge step 1}
\label{tabivy1}
\begin{tabular}{|l|l|l|l|l|l|}
\hline
&0.5&2&5&10&20\\
\hline
$R_i$&99.96&100.17&101.16&103.66&108.57\\
\hline
$R_v$&89.3&111.16&126.05&137.35&148.66\\
\hline
$R_i$ \& $R_v$&89&111.18&126.06&137.36&148.67\\
\hline
\end{tabular}
\vskip1pt
\caption{Ivy Bridge step 2}
\label{tabivy2}
\begin{tabular}{|l|l|l|l|l|l|}
\hline
&1.1&1.25&1.5&2\\
\hline
"cmd r 1","cmd x 1"&100&100&100&100\\ 
\hline
"cmd r,r 1"&100.013&100.035&100.072&100.144\\
\hline
"cmd x,x 1"&100.011&100.0297&100.058&100.117\\
\hline
"cmd r,r,r 1"&100&100.431&101.049&101.945\\
\hline
"cmd x,x,x 1"&100&100.334&100.799&101.496\\
\hline
\end{tabular}
\vskip1pt
\caption{Ivy Bridge step 3}
\label{tabivy3}
\begin{tabular}{|l|l|l|l|l|l|}
\hline
&8&16&32&64\\
\hline
$r \times 2$ cmd3&136.298&140.419&144.358&148.213\\
\hline
$x \times 2$ cmd4&134.032&138.137&142.07&145.921\\
\hline
$r \times 5$ cmd3&156.223&160.361&164.309&168.168\\
\hline
$x \times 5$ cmd4&153.939&158.072&162.016&165.874\\
\hline
$r \times 10$ cmd3&171.31&175.454&179.404&183.265\\
\hline
$x \times 10$ cmd4&169.02&173.162&177.111&180.97\\
\hline
\end{tabular}
\end{table}

\subsection{Haswell and Skylake}
Haswell processors (tables \ref{tabhas1},\ref{tabhas2} and \ref{tabhas3}) were improved by increasing the number of registers and by adding some instructions of the existing types. We can observe that most of investigated characteristics are unchanged for Ivy Bridge, Haswell and Skylake. The main tendencies are for increasing the number of registers (but not as much as for Intel Core), changing the instruction set and making some improvements to the processor pipeline. 

\begin{table}[!ht]
\centering
\caption{Haswell step 1}\label{tabhas1}
\begin{tabular}{|l|l|l|l|l|l|}
\hline
&0.5&2&5&10&20\\
\hline
$R_i$&99.71&101.85&110.89&120.75&131.04\\
\hline
$R_v$&91.57&110.03&123.66&134.01&144.37\\
\hline
$R_i$ \& $R_v$&89.72&110.32&123.99&134.34&144.7\\
\hline
\end{tabular}
\vskip1pt
\caption{Haswell step 2}\label{tabhas2}
\begin{tabular}{|l|l|l|l|l|}
\hline
&1.1&1.25&1.5&2\\
\hline
"cmd r,r 1","cmd x,x 1"&100.003&100.008&100.015&100.03\\
\hline
"cmd r,r,r 1","cmd x,x,x 1"&100.177&100.473&100.904&101.669\\
\hline
\end{tabular}
\vskip1pt
\caption{Haswell step 3}\label{tabhas3}
\begin{tabular}{|l|l|l|l|l|}
\hline
&8&16&32&64\\
\hline
$r \times 2$ cmd3, $x \times 2$ cmd4&125.817&129.492&133.053&136.558\\
\hline
$r \times 5$ cmd3, $x \times 5$ cmd4&143.936&147.683&151.277&154.798\\
\hline
$r \times 10$ cmd3, $x \times 10$ cmd4&157.7&161.471&165.076&168.603\\
\hline
\end{tabular}
\end{table}

\subsection{Skylake and the prediction of its successor}
The results obtained for this last processor from our list (tables \ref{tabsky1},\ref{tabsky2} and \ref{tabsky3}) present the direction for making predictions for succeeding processors. We can note that as far as the characteristics from the technical group are concerned, the biggest effect on the computer capacity is reached by increasing the number of vector registers. To achieve the effect close to the Intel Core -- Ivy Bridge capacity increase we need to increase tenfold the number of registers and to add a large number of new fast instructions with four registers. 

\begin{table}[!ht]
\centering
\caption{Skylake step 1}\label{tabsky1}
\begin{tabular}{|l|l|l|l|l|l|}
\hline
&0.5&2&5&10&20\\
\hline
$R_i$&99.66&102.08&111.51&121.41&131.67\\
\hline
$R_v$&91.83&109.95&123.53&133.85&144.17\\
\hline
$R_i$ \& $R_v$&89.76&110.29&123.92&134.24&144.56\\
\hline
\end{tabular}
\caption{Skylake step 2}\label{tabsky2}
\begin{tabular}{|l|l|l|l|l|}
\hline
&1.1&1.25&1.5&2\\
\hline
"cmd r,r 1"&100.004&100.01&100.019&100.039\\
\hline
"cmd x,x 1"&100.003&100.008&100.017&100.034\\
\hline
"cmd r,r,r 1"&100.201&100.585&101.111&102.021\\
\hline
"cmd x,x,x 1"&100.166&100.483&100.924&101.702\\
\hline
\end{tabular}
\caption{Skylake step 3}\label{tabsky3}
\begin{tabular}{|l|l|l|l|l|}
\hline
&8&16&32&64\\
\hline
$r \times 2$ cmd3&126.767&130.454&134.014&137.514\\
\hline
$x \times 2$ cmd4&125.454&129.111&132.657&136.151\\
\hline
$r \times 5$ cmd3&144.858&148.603&152.19&155.703\\
\hline
$x \times 5$ cmd4&143.511&147.243&150.825&154.335\\
\hline
$r \times 10$ cmd3&158.589&162.354&165.95&169.467\\
\hline
$x \times 10$ cmd4&157.23&160.989&164.582&168.098\\
\hline
\end{tabular}
\end{table}

The first step of our research shows that some parameters  do not affect the performance starting from the first examined processor. These parameters are sizes of all types of memory and theirs latencies. Certainly, we do not want to claim that these parameters are useless and can be freely removed, but in the previous models the level of saturation was reached for these parameters, so increasing them more is inefficient. Also we isolate the parameters which have a significant influence on the value of computer capacity and that have been changed substantially during the evolution process. In fact, the registers of a processor is the fastest memory (they can be accessed nearly instantly), so it is expected that this value has the highest influence among all the presented parameters. Nevertheless, it is important to note that the really significant effect on the value of computer capacity (more than 10\%) is obtained at 10-20 fold increase. 

The second step of research shows the influence of the instruction set on the performance of processors. All the obtained results show the growth of the computer capacity in the range of 1-2\% even with a double increase of the number of fastest instructions. Considering the complexity of designing new instructions, we can speak of the saturation of the instruction set with the existing types of instructions. It is also clearly seen that this exact way has been chosen by the manufacturers of the processors. Starting from Ivy Bridge processor a new type of fast instructions with three register operands was added. This statement is proved 
 by the results from the third step of research where we simultaneously increase the number of registers and add instructions with three operands. As we can see in Table \ref{tabops}, 10 instructions with three register operands were added, and in Table \ref{tabSum} we see that the number of integer registers increased from 16 to 160 (10 times). 
 The corresponding predicted value in  Table \ref{tabcore3} is 147.49\%, which is close to  
 the original values of computer capacity for processors Intel Core and Ivy Bridge from table \ref{tabSum}, which is 153.16\%.

\section{Conclusions}\label{sec3}

After presenting all the obtained results we can make some conclusions. The last considered microarchitecture is Skylake (Kaby lake is just its
modification) and its successor Cannon Lake, whose characteristics we try to predict,  is not released yet. The investigations show that just a 
small set of Skylake parameters affects the computer capacity. In order to achieve greater performance the manufacturer needs to increase the number 
of registers (as we can see in table \ref{tabsky1}, vector registers have a stronger effect on the computer capacity value then integer registers) 
and add some new processor instructions. On the other hand, it is clearly seen that it makes no sense to change the values of RAM and cache-memory sizes 
or access times. In the context of presented work we do not consider the other parameters whose influence is obvious and linear (the number of cores, the clock  rate etc.). So, we can predict that in Cannon Lake the manufacturer will increase the number of registers and add some new vector instructions with three or more operands. 

We can see that a direct application of computer capacity gives a new approach to analyse computer performance. 
Generally speaking, all changes are aimed at increasing computing capacity, defined in (\ref{eq1}) (and estimated based on the equation (\ref{eqCC}) )
In this paper we successfully applied 
the computer capacity for investigation and prediction of processor evolution.

\appendix
\section*{APPENDIX}
\begin{figure}[!ht]
	\includegraphics[width=\textwidth]{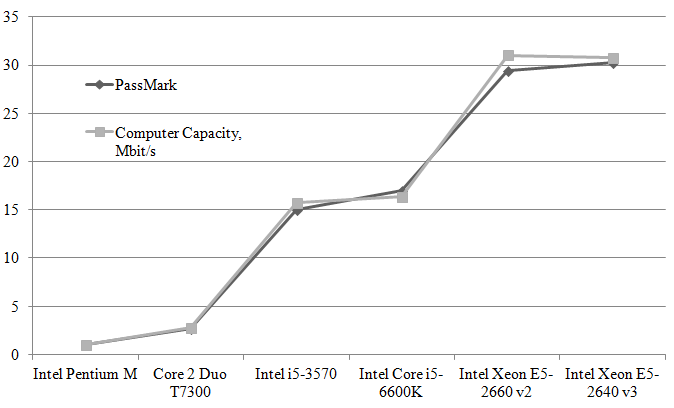}
	\caption{The results of processors comparison}
	\label{fig1}
\end{figure}

\setcounter{section}{1}
\label{appendix}
In table \ref{tabcomp} we present the values of Computer Capacity and PassMark benchmark for all processor microarchitectures investigated in paper. Here, the processor Ivy Bridge was presented twice just to show how Computer Capacity would behave with different numbers of cores (i5-3579 has 4 cores and E5-2660v2 has 10 cores). At figure \ref{fig1} we can see the results of comparison these values. As the measurement units are different we took the value of the first processor in list (Intel Pentium M) as 1 and build the graph in relation to the values of this processor. For example, we divide the value of Computer Capacity for i5-6600K (1673398.77) by 102434.84 and get the result 16.336, and analogously we divide the value of PassMark benchmark (7884) by 464 and get 16.991. 
\begin{table}[!ht]
\centering
\caption{The values of Computer capacity and PassMark benchmark}\label{tabcomp}
\begin{tabular}{|l|l|l|}
\hline
Name&PassMark&Computer \\
&&capacity, Mbit/s\\
\hline
Intel Pentium M&464&102434.84\\
\hline
Core 2 Duo T7300 (Intel Core)&1232&283593.72\\
\hline
Intel i5-3570 (Ivy Bridge)&6978&1610797.70\\
\hline
Intel Core i5-6600K (Skylake)&7884&1673398.77\\
\hline
Intel Xeon E5-2660v2 (Ivy Bridge)&13659&3179205.98\\
\hline
Intel Xeon E5-2640v3(Haswell)&14036&3151395.48\\
\hline
\end{tabular}
\end{table}

\newpage


\begin{thebibliography}{10}
\providecommand{\url}[1]{#1}
\csname url@samestyle\endcsname
\providecommand{\newblock}{\relax}
\providecommand{\bibinfo}[2]{#2}
\providecommand{\BIBentrySTDinterwordspacing}{\spaceskip=0pt\relax}
\providecommand{\BIBentryALTinterwordstretchfactor}{4}
\providecommand{\BIBentryALTinterwordspacing}{\spaceskip=\fontdimen2\font plus
\BIBentryALTinterwordstretchfactor\fontdimen3\font minus
  \fontdimen4\font\relax}
\providecommand{\BIBforeignlanguage}[2]{{%
\expandafter\ifx\csname l@#1\endcsname\relax
\typeout{** WARNING: IEEEtran.bst: No hyphenation pattern has been}%
\typeout{** loaded for the language `#1'. Using the pattern for}%
\typeout{** the default language instead.}%
\else
\language=\csname l@#1\endcsname
\fi
#2}}
\providecommand{\BIBdecl}{\relax}
\BIBdecl

\bibitem{perf_eval}
B.~Ryabko, ``An information-theoretic approach to estimate the capacity of
  processing units,'' \emph{Performance Evaluation}, vol.~69, pp. 267--273,
  2012.

\bibitem{jcsc}
B.~Ryabko and A.~Rakitskiy, ``An analytic method for estimating the computation
  capacity of computing devices,'' \emph{Journal of Circuits, Systems and
  Computers}, vol.~26, no.~05, p. 1750086, 2017.

\bibitem{weicker1991detailed}
R.~P. Weicker, ``A detailed look at some popular benchmarks,'' \emph{Parallel
  Computing}, vol.~17, no. 10-11, pp. 1153--1172, 1991.

\bibitem{overall}
D.~J. Lilja, \emph{Measuring computer performance: a practitioner's
  guide}.\hskip 1em plus 0.5em minus 0.4em\relax Cambridge university press,
  2005.

\bibitem{passmark}
\BIBentryALTinterwordspacing
PassMark\_Software, ``Passmark performance test description,'' 2016. [Online].
  Available:
  \url{http://www.passmark.com/support/performancetest/interpreting\_test\_results.htm}
\BIBentrySTDinterwordspacing

\bibitem{red16}
A.~Rakitskiy and B.~Ryabko, ``An information-theoretic approach to performance
  evaluation of supercomputers,'' in \emph{2016 XV International Symposium
  Problems of Redundancy in Information and Control Systems (REDUNDANCY)}, Sept
  2016, pp. 125--128.

\bibitem{shannon}
C.~E. Shannon, ``A mathematical theory of communication,'' \emph{Bell system
  technical journal}, vol.~27, 1948.

\bibitem{phansalkar2005measuring}
A.~Phansalkar, A.~Joshi, L.~Eeckhout, and L.~K. John, ``Measuring program
  similarity: Experiments with spec cpu benchmark suites,'' in
  \emph{Performance Analysis of Systems and Software, 2005. ISPASS 2005. IEEE
  International Symposium on}.\hskip 1em plus 0.5em minus 0.4em\relax IEEE,
  2005, pp. 10--20.


\bibitem{agner1}
\BIBentryALTinterwordspacing
A.~Fog, ``The microarchitecture of intel, amd and via cpus,'' 2016. [Online].
  Available: \url{www.agner.org/optimize/}
\BIBentrySTDinterwordspacing

\bibitem{link}
\BIBentryALTinterwordspacing
A.~Rakitskiy, ``Characteristic equations and evaluation programs,'' 2016.
  [Online]. Available:
  \url{http://www.ict.nsc.ru/ru/structure/orgunits/lab-info-sys-security-page}
\BIBentrySTDinterwordspacing

\end{thebibliography}



\appendix
\section*{APPENDIX 2}\label{appendix2}
\medskip
The characteristic equation (\ref{eqCC}) for Haswell processor:
\begin{eqnarray*}
\frac{ 524014680 }{X^{ 1 }} + \frac{ 2448358122 }{X^{ 2 }} + \frac{ 81258594 }{X^{ 3 }} + \frac{ 4863059 }{X^{ 4 }} + \frac{ 19968745476 }{X^{ 5 }} + \frac{ 7936385025 }{X^{ 6 }} + \\
\frac{ 118510993664 }{X^{ 7 }} + \frac{ 4187392636 }{X^{ 8 }} + \frac{ 4198458 }{X^{ 9 }} + \frac{ 232644384 }{X^{ 10 }} + \frac{ 6978 }{X^{ 11 }} + \frac{ 14300162 }{X^{ 12 }} + \\
\frac{ 3440640 }{X^{ 13 }} + \frac{ 19117056 }{X^{ 14 }} + \frac{ 9516033 }{X^{ 15 }} +\frac{ 32768 }{X^{ 16 }} + \frac{ 79874982059 }{X^{ 17 }} + \frac{ 31745540097 }{X^{ 18 }} + \\
\frac{ 474046726144 }{X^{ 19 }} + \frac{ 16768421479 }{X^{ 20 }} +\frac{ 16678913 }{X^{ 21 }} + \frac{ 949315584 }{X^{ 22 }} + \frac{ 2778726 }{X^{ 23 }} + \frac{ 19267584 }{X^{ 24 }} + \\
\frac{ 13762561 }{X^{ 25 }} +\frac{ 39911425 }{X^{ 26 }} + \frac{ 131072 }{X^{ 27 }} + \frac{ 2752513 }{X^{ 28 }} + \frac{ 11010049 }{X^{ 31 }} + \frac{ 3145728 }{X^{ 288 }} + \\
\frac{ 9483265 }{X^{ 34 }} + \frac{ 11010048 }{X^{ 35 }} + \frac{ 168 }{X^{ 36 }} + \frac{ 5505024 }{X^{ 38 }} + \frac{ 11010048 }{X^{ 40 }} +\frac{ 1 }{X^{ 41 }} + \frac{ 6553 }{X^{ 47 }} + \\
\frac{ 3932160 }{X^{ 50 }} + \frac{ 9584997826560 }{X^{ 51 }} + \frac{ 3809464811520 }{X^{ 52 }} + \frac{ 56885276835840 }{X^{ 53 }} + \\
\frac{ 2009934594048 }{X^{ 54 }} + \frac{ 2001469442 }{X^{ 55 }} + \frac{ 111641886720 }{X^{ 56 }} + \frac{ 3145728 }{X^{ 57 }} + \\
\frac{ 2312110081 }{X^{ 58 }} + \frac{ 1651533582 }{X^{ 59 }} + \frac{ 4624220160 }{X^{ 60 }} + \frac{ 15728640 }{X^{ 61 }} + \\
\frac{ 1717986918 }{X^{ 318 }} + \frac{ 1321205760 }{X^{ 65 }} + \frac{ 1321205760 }{X^{ 69 }} + \frac{ 1 }{X^{ 70 }} + \frac{ 1 }{X^{ 71 }} + \frac{ 660602880 }{X^{ 72 }} + \frac{ 1321205760 }{X^{ 74 }} + \\
\frac{ 1 }{X^{ 78 }} + \frac{ 2147483648 }{X^{ 80 }} + \frac{ 5234686813011968 }{X^{ 81 }} + \frac{ 2080475715731456 }{X^{ 82 }} +\\
\frac{ 31066945855946752 }{X^{ 83 }} + \frac{ 1097692279629414 }{X^{ 84 }} + \frac{ 1093069176832 }{X^{ 85 }} + \frac{ 60971355734016 }{X^{ 86 }} + \\
\frac{ 1717986918 }{X^{ 87 }} +\frac{ 1262720385024 }{X^{ 88 }} + \frac{ 901943132160 }{X^{ 89 }} + \frac{ 2525440770048 }{X^{ 90 }} + \\
\frac{ 8589934592 }{X^{ 91 }} + \frac{ 3145728 }{X^{ 93 }} + \frac{ 8192 }{X^{ 94 }} + \frac{ 721554505728 }{X^{ 95 }} + \frac{ 721554505728 }{X^{ 99 }} + \frac{ 360777252864 }{X^{ 102 }} + \\
\frac{ 721554505728 }{X^{ 104 }} + \frac{ 32768 }{X^{ 106 }} + \frac{ 2 }{X^{ 110 }} + \frac{ 128 }{X^{ 120 }} + \frac{ 1717986918 }{X^{ 123 }} + \frac{ 512 }{X^{ 132 }} + \frac{ 128 }{X^{ 134 }} + \\
\frac{ 3932160 }{X^{ 140 }} + \frac{ 512 }{X^{ 146 }} + \frac{ 8192 }{X^{ 151 }} + \frac{ 32768 }{X^{ 163 }} + \frac{ 61440 }{X^{ 166 }} + \frac{ 2147483648 }{X^{ 170 }} + \frac{ 1 }{X^{ 173 }} + \frac{ 61440 }{X^{ 180 }} + \\
\frac{ 33554432 }{X^{ 196 }} + \frac{ 3932160 }{X^{ 197 }} + \frac{ 33554432 }{X^{ 210 }} + \frac{ 1 }{X^{ 224 }} + \frac{ 2147483648 }{X^{ 227 }} + \frac{ 6553 }{X^{ 242 }} + \frac{ 26214 }{X^{ 254 }} = 1
\end{eqnarray*}

\end{document}